\documentclass[pre,twocolumn,showpacs]{revtex4}
\usepackage{graphicx}
\usepackage{amsmath}
\usepackage{amsfonts}
\usepackage{mathrsfs}
\newcommand{\kbar}{\mathchar'26\mkern-9muk}
\begin{document}
\title{Observation of Robust Quantum Resonance Peaks in an Atom Optics Kicked Rotor with Amplitude Noise}
\author{Mark Sadgrove}
\author{Andrew Hilliard}
\author{Terry Mullins}
\author{Scott Parkins}
\author{Rainer Leonhardt}
\affiliation{Department of Physics, The University of Auckland, Private Bag 92019,
Auckland, New Zealand}
\begin{abstract}
The effect of pulse train noise on the quantum resonance peaks of
the Atom Optics Kicked Rotor is investigated experimentally. Quantum 
resonance peaks in the late time mean energy of the atoms are found to be 
surprisingly robust against all levels of noise applied to the kicking amplitude, whilst
even small levels
of noise on the kicking period lead to their destruction.
The robustness to amplitude noise of the resonance peak and of the fall--off in
mean energy to either side of this peak are explained in terms of the occurence of
stable, $\epsilon$--classical dynamics [S. Wimberger, I. Guarneri, and S. Fishman,
\textit{Nonlin.} \textbf{16}, 1381 (2003)] around each quantum resonance.
\end{abstract}

\maketitle

\section{Introduction}
The sensitivity of coherent quantum phenomena 
to the introduction of extraneous degrees of freedom is well documented \cite{Bundle1}. 
In particular, the coupling of a quantum system to its 
environment or, equivalently, subjection of the system to measurement
is known to result in decoherence, that is, a loss of quantum interference phenomena. 

The experimental study of decoherence ideally requires a system whose 
coupling to the environment may be completely 
controlled. The discipline of Atom Optics allows the realisation of 
this requirement in the form of atoms interacting with a far detuned 
optical field. The Atom Optics Kicked Rotor (AOKR),
first implemented experimentally by the Raizen Group of Austin, Texas
\cite{Moore1995,Raizen1996}, is a particular example of some 
interest as it is a quantum system which is chaotic in the classical
limit. The AOKR is realised by subjecting cold atoms to short, periodic
pulses of an optical standing wave detuned from atomic resonance.
The atoms typically experience curtailed energy growth (dynamical
localisation) \cite{Fishman1982} compared with the classical case, but may also experience
enhanced growth for certain pulsing periods, an effect known as
\emph{quantum resonance} \cite{Izrailev1979}.

Previous AOKR experiments have shown that spontaneous emission events
and noise applied to the amplitude of the kicking pulse train result 
in the destruction of quantum dynamical localisation 
\cite{NelsonPRL,Raizen1998a}. It might 
then be expected that the other well known signature of quantum dynamics in 
the AOKR,
quantum resonance, should exhibit great sensitivity to spontaneous emission
or noisy pulse trains. However, recent experiments by d'Arcy \emph{et al.}
have shown that detection of quantum resonance behaviour is actually
enhanced in the presence of spontaneous emission \cite{dArcy2001, dArcy2001b, dArcy2003pp},
in stark contrast to the accepted wisdom on the effects of spontaneous emission noise.
Recent numerical work has also focussed on the susceptibility of quantum resonance
behaviour to applied noise \cite{Brouard2003}.

Here we present further experimental evidence of the robustness of
the quantum resonance peak to certain types of noise. In this case, 
noise is added to the kicked rotor system by introducing random
fluctuations in the amplitude or period of the optical pulses used
to kick the atoms (collectively termed \emph{pulse train noise}).
We find that even in the presence of maximal amplitude noise,
the structure near to quantum resonance persists
(including the low energy levels to either side of the peak).
This resistance to amplitude fluctuations 
runs counter to the expectation that quantum phenomena are sensitive to
noise. In contrast, a small amount of noise added to the period of the 
pulses is enough to completely wash out the resonance structure.
The robustness of the near resonant behaviour to amplitude noise is 
reminiscent of recent observations of quantum stability in the quantum
kicked accelerator by Schlunk \emph{et al.} \cite{Schlunk2003a,Schlunk2003b}.

The remainder of this paper is arranged as follows:
Section \ref{sec:AOKR} provides background on the formal 
AOKR system with amplitude and period noise. Section 
\ref{sec:Noise} reviews the study of quantum resonances in the
kicked rotor. Our experimental procedure and results are
found in Sections \ref{sec:ExpSet} and \ref{sec:ExpRes} respectively,
and the results are explained in Section \ref{sec:EpsClass} in terms of the 
recently developed $\epsilon$--classical model for quantum resonance peaks.
Section\ref{sec:Conclusion} offers conclusory remarks.

\section{Atom optics kicked rotor with amplitude and period noise}
\label{sec:AOKR}

The Hamiltonian for an AOKR kicked with period $T$ with fluctuations 
in the amplitude and/or pulse timing is given in scaled units by
\begin{equation}
\label{ham1} \hat{H}=\frac{\hat{\rho}^2}{2}-\kappa
\cos(\hat{\phi})\sum^{N}_{n=0}R_{A,n}f(\tau-n R_{P,n}),
\end{equation}
where $\hat{\phi}$ and $\hat{\rho}$ are the quantum operators for the
(scaled) atomic position and momentum, respectively, $\kappa$ is the
kicking strength, $f$ is the pulse shape function,
 $\tau = t/T$ is the scaled time and the terms $R_{A,n}$ and $R_{P,n}$
introduce random fluctuations in the amplitude and kicking
period respectively. We also note the scaled commutator relationship
$[\hat{\phi}, \hat{\rho}] = i\kbar$, where $\kbar=8\omega_rT$ is a scaled Planck's
constant and $\omega_r$ is the frequency associated with the energy change after a single
photon recoil for Caesium. The scaled momentum $\hat{\rho}$ is related to the atomic momentum 
$\hat{p}$ by the equation $\hat{\rho}/\kbar = \hat{p}/(2 \hbar k_L)$, where $k_L$ is the wave number 
of the laser light.
In this paper, as in Refs. \cite{Auckland2003, dArcy2001}, momentum is presented in the
``experimental units'' of $\hat{p}/(2 \hbar k_L)$.

Assuming, for simplicity, a rectangular pulse shape, 
the stochasticity parameter $\kappa$ is related to experimental parameters
by the equation 
\begin{equation}
\kappa = \Omega_{\rm eff}\omega_rT\tau_p,
\label{Eq:Omega_eff}
\end{equation}
where $\Omega_{\rm eff}$ is the potential strength created by the laser field,
and $\tau_p$ is the duration of the kicking pulse. $\Omega_{\rm eff}$
is given by
\begin{equation} 
\Omega_{{\rm eff}}=\frac{\Omega^2}{\Delta},
\label{Eq:Omega_eff2}
\end{equation} 
where $\Omega$ is the resonant, single beam Rabi frequency of the atoms 
and $\Delta$ (which is $\approx 2\times10^9$ rad ${\rm s}^{-1}$ for
these experiments) takes into account the relative transition strengths between and 
laser detunings from the different hyperfine states of caesium,
as discussed in our previous papers (see, for example, 
\cite{NelsonPRL,Auckland2003}).

Noise is introduced by the terms $R_{i,n} = 1+\delta_{i,n}$, 
where $\delta_{i,n}$ is a random variable with probability distribution
\begin{equation}
P(\delta_{i,n}) = \left\{ 
\begin{array}{ll}
1/{\rm \mathcal{L}_i}, & |\delta_{i,n}| < {\mathcal{L}_i/2}\\
0, & \rm{else}
\end{array}
\right.
\end{equation}
with $i=A$ denoting amplitude noise, and $i=P$ denoting period noise.
The noise level is denoted $\mathcal{L}_i$. For amplitude noise,
we have $0 \leq {\mathcal{L}_{A}} \leq 2$, where a noise level of $2$
corresponds to the case where the kicking strength can vary between 
$0$ and $2\overline{\kappa}$ for each pulse, with $\overline{\kappa}$ 
the mean value of the kicking strength in the experiment.
For period noise, $0 \leq {\mathcal{L}_{P}} < {\mathcal{L}_{P, {\rm max}}}$ where
${\mathcal{L}_{P, {\rm max}}}$ is $1$ for the $\delta$-kicked rotor and 
$1-\alpha$ for the pulse kicked rotor in our experiments, with
$\alpha$ the ratio of the pulse width to the pulse period (less than $1\%$ in our experiments).
We note that our implementation of period noise differs from that used in
\cite{RaizenPeriod} in that it shifts each pulse a random amount from its
zero--noise position rather than randomising the timing between consecutive
pulses. This means that the effect of the period noise fluctuations
is not cumulative (as it is in the aforementioned reference), allowing 
a more instructive comparison of the effects of period noise with those 
of amplitude noise.

\section{The quantum resonances of the AOKR}
\label{sec:Noise}
In a fully chaotic driven system no stable periodic orbits exist in phase space and thus no 
frequency of the driving force gives rise to resonant behaviour. Although the
classical kicked rotor retains kick--to--kick correlations for any value
of the stochasticity parameter, for sufficiently high $\kappa$ the phase space is 
essentially chaotic, and the dynamics are independent of the kicking period of the system.
However, this is not true of the quantum system, even for large $\kappa$,
as fundamental periodicities exist in the quantum dynamics. This may be seen
by inspecting the one kick evolution operator for the quantum $\delta$--kicked 
rotor, which has the form 
\begin{equation}
\hat{U} = \exp({\rm i}\kappa \cos\hat{\phi}/\kbar)\exp(-{\rm i}\hat{\rho}^2/(2\kbar)),
\label{eq:evolrotor}
\end{equation} 
For the analysis of quantum resonance, the second 
exponential term (or \emph{free evolution} term) of Eq. \ref{eq:evolrotor}
is of primary importance. We see that if $\kbar$ is an even multiple of $2\pi$, and the state
undergoing evolution is a momentum eigenstate (or a quantum superposition of 
such eigenstates) $|n\rangle$ such that
$\hat{\rho}|n\rangle = n\kbar|n\rangle$, this term becomes unity. This is the quantum resonance
condition, and it may be shown that atoms initially in momentum eigenstates undergo
ballistic motion \cite{Izrailev1979} at resonant values of $\kbar$.
For $\kbar=2\pi(2m-1)$, $m$ a positive integer, initial 
momentum eigenstates with even and odd $n$ acquire quantum phases after free evolution 
of $+1$ and $-1$ respectively. It is found that the additional possibility of $-1$ for
the phase of odd momentum components of the wavefunction 
leads to oscillations in the mean energy of the kicked atomic ensemble
 \cite{Oskay2000, Deng1999}. Thus, $\kbar=2\pi$ is termed a \textit{quantum antiresonance}.

We note that whilst quantum resonances are predicted to exist for all
rational multiples of $\kbar=2\pi$, resonance peaks have only been observed
in experiments and simulations at integral multiples. In this paper, we 
focus on the behaviour at $\kbar=2\pi$ and $\kbar=4\pi$ and will refer to 
the energy peaks at these values of the scaled Planck's constant as the 
first and second quantum resonances respectively.

For a cloud of Caesium atoms at $5\mu$K, as used in our experiments,
the atomic momentum distribution has a standard deviation of 
$\sim 5 \hbar k_L$, so only a small momentum subclass of the atoms 
may be considered to be in an initial momentum eigenstate. In general,
each atom has a momentum of the form $\rho = n + \beta$ (in scaled units), where $n$ is an 
integer and $\beta \in [0,1)$ is known as a \emph{quasimomentum}.
The appropriate evolution operator when the quasimomentum of the 
atoms is included is
\begin{equation}
\hat{U}_\beta = \exp({\rm i}\kappa \cos\hat{\phi}/\kbar)\exp(-{\rm i}(\hat{n}+\beta)^2/(2\kbar)).
\label{eq:evolbeta}
\end{equation}
For some values of quasimomenta, this one--kick evolution operator still exhibits the 
periodicity necessary for resonance \cite{Wimberger2003}. Specifically, ballistic energy
growth occurs for $\kbar=2\pi$, $\beta=0.5$ and for $\kbar=4\pi$, $\beta=0$ or $0.5$.

The quantum resonances of the AOKR were first studied experimentally
by the group of Mark Raizen at Austin, Texas \cite{Moore1995, 
Raizen1996,Oskay2000}. In particular, ref. \cite{Moore1995} presented the
results of experiments in which the momentum distribution of the atoms
was recorded for various kicking periods. The momentum distributions corresponding
to quantum resonance were found to be \textit{narrower} than those off resonance. The relatively small 
population of atoms undergoing ballistic energy growth
at resonance was not detected experimentally and no difference was found
between momentum distributions for odd and even multiples of $\kbar=2\pi$.
In ref. \cite{Oskay2000} a further
study by the group detected the expected ballistic peaks at $\kbar=2\pi$
and $\kbar=4\pi$. Additionally, small oscillations in the widths of 
the atomic momentum distributions as a function of kick number 
were seen only at $\kbar=2\pi$ -- a result of the anti-resonance behaviour described earlier.

More recent experiments by d'Arcy \emph{et al.} \cite{dArcy2001,dArcy2001b,
dArcy2003pp} have focussed on the
effect of spontaneous emission on the quantum resonance peaks.
They found experimentally that spontaneous emission makes these peaks more 
prominent -- a somewhat counter--intuitive result. Further theoretical 
investigations revealed that this 
phenomenon was due to the reshuffling of atomic quasimomenta caused by
spontaneous emission which allows more atoms to experience resonant behaviour 
at some time during their evolution. Additionally, reshuffling of quasimomenta
results in fewer atoms gaining large momenta from multiple resonant
kicks. Without spontaneous emission, resonant atoms  soon travel outside 
the finite observation window of the experiment and thus do not contribute
to the measured energy of the atomic ensemble.

Our experiments measure the structure of the mean energy around the quantum 
resonance peak in a similar
fashion to the experiments described above. The pulse period is scanned over
the resonant value and the mean energy is extracted from the measured momentum
distributions at each value of $T$. For the power and detuning of the kicking laser 
used in this experiment, there is a constant chance of spontaneous
emission per pulse of $\sim 2.5\%$. As in \cite{dArcy2001}, this is found
to increase the height and width of the resonance peaks and make them more amenable to
investigation. Our numerical studies show that the non-zero spontaneous emission
rate does not affect the study of amplitude noise and period noise
on the quantum resonance peak. This is because the mechanisms by which
pulse train noise and spontaneous emission noise influence the atomic dynamics
are totally different: Spontaneous emission
events affect individual atoms by changing their quasimomenta; Amplitude
and period noise change the kick--to--kick correlations over the entire
atomic ensemble and do not change atomic quasimomenta. Thus the
advantages of a relatively high spontaneous emission rate may be utilised
without biasing the study of the effects of pulse train noise on the quantum resonance
peaks.

\section{Experimental setup}
\label{sec:ExpSet}
Our experiments utilise a $5$ $\mu$K cloud of cold Caesium atoms, provided 
by a standard six beam magneto--optical trap (MOT) \cite{Monroe1990}. The atoms interact 
with a pulsed, far-detuned optical standing wave which is created by 
retroreflecting the light from a 150mW (slave) diode laser which is
injection locked to a lower power (master) diode laser. The output of the master laser may be tuned
over a range of about $4$ GHz relative to the $6S_{1/2}(F=4) \rightarrow 6P_{3/2}(F'=5)$ transition
of the Caesium $\rm D_2$ line. The detuning of the laser from this transition is denoted $\delta$.
The frequency of the kicking laser is monitored by observing the spectrum of its beat signal
with the trapping laser.

The standing wave has a horizontal orientation rather than the vertical
orientation used in the quantum accelerator experiments of references 
\cite{Schlunk2003a, Schlunk2003b}.
It is pulsed by optically switching the laser light using an 
acousto--optic modulator (AOM). The amplitude of the AOM's driving signal
is controlled by a programmable pulse generator (PPG) to achieve the desired
pulse train shape. For amplitude noise experiments, the AOM's
response to the amplitude of its driving signal must first be calibrated,
since the pulse heights need to be uniformly distributed.
The PPG consists of a random access memory (RAM) chip which can store up
to $2^{16}$ $12$ bit words representing samples of the pulse train. On 
receipt of a gate pulse, the samples in the RAM are read into a digital 
to analogue convertor at 25 MHz,
corresponding to a 40ns temporal resolution for the pulse trains. 
A given realisation of a noisy pulse train (for amplitude or period noise)
is created by using computer--generated pseudo--random numbers to give
fluctuations about the mean amplitude or mean pulse position in a standard
pulse train. The noisy pulse train is then uploaded to the PPG. 

In a typical experimental run, the cooled atoms were released from the 
MOT and subjected to $20$ standing wave pulses, then allowed to
expand for an additional free drift time in order to resolve
 the atomic momenta. The momentum resolution of our
experiments for a $12$ms expansion time is $0.29$ $2$--photon recoils.
After free expansion, the atoms were subjected
to optical molasses, effectively freezing them in place, and a 
fluoresence image of their spatial distribution was taken.  
The timing of the experiment was controlled by sequencing software
running on the $\rm RTLinux^{\tiny \texttrademark}$ operating system kernel
giving worst case timing errors of $30$ $\mu$s \cite{RTLinuxFAQ}, or $0.25\%$ 
of the atomic time of flight.

Some experimental imperfections have a systematic effect on our
data and need to be taken into account in simulations in order
for meaningful comparisons to be made. Firstly, when the standing wave
is on, individual atoms experience differing potentials depending on their
radial position in the beam, due to the gaussian mode shape of the beam.
This can affect the experimental resolution of the multi--peaked `diffusion resonance'
structure in the mean energy which occurs between primary quantum resonances, as this structure
is strongly dependent on the exact potential strength \cite{Auckland2003,Daley2001}.
However it is not so critical to the observation of quantum resonance peaks, due to the
very resistance to variations in amplitude discussed in this paper.
Nonetheless, this spread in kicking strengths is taken into 
account in our simulations. Secondly, in order to achieve a spontaneous 
emission rate sufficiently high to make the quantum resonance peaks prominent
and amenable to study,
a detuning from resonance of about $500$ MHz was used in our experiments.
This value of the detuning is large enough to ensure the condition $\delta \gg \overline{\Omega}$
(where $\overline{\Omega}$ is the average atomic Rabi frequency taken over the different hyperfine transitions) which is assumed
in the derivation of the AOKR Hamiltonian \cite{Graham1992}.
However, the difference in detuning between the $F=4$ 
ground state and each of the hyperfine excited states $F' = 3,4,5$, 
as well as the difference in coupling strengths between magnetic substates,
leads again to a spread in kicking strengths (as detailed in reference 
\cite{Auckland2003}). Once again, this effect is allowed for in our 
simulations.

We also note that the application of amplitude and period noise to our pulse trains 
inherently creates random scatter in our data since each different 
noise realisation gives rise to a different mean energy. Thus, meaningful results
may only be obtained by averaging the energy from a number of separate experiments with 
different noise realisations. For experiments where the noise is solely
a result of spontaneous emission events, the statistics are already excellent, since
the mean energy is calculated for a large number of individual atoms. This is not true for pulse
train noise experiments which affect correlations over the entire atomic ensemble. 
Each point on our curves represents an average of 12 separate experiments (except in the
zero noise case, where 3 repetitions was found to be sufficient).
This number of repetitions reduces the error to a size such that any quantum 
resonance structure may be confidently identified.

\section{Experimental and simulation results}
\label{sec:ExpRes}
\begin{figure}[thb]
\includegraphics[height=7cm]{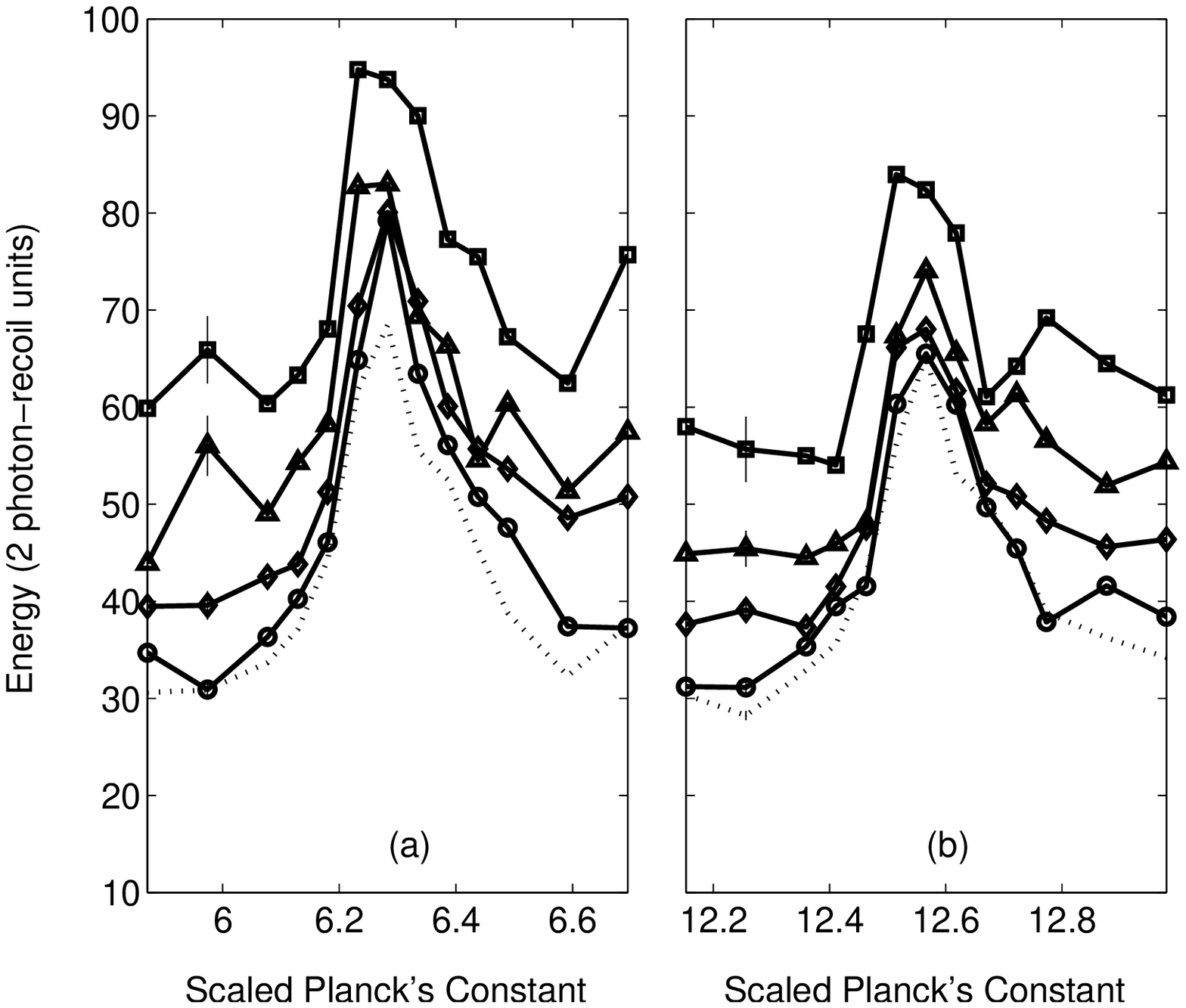}
\includegraphics[height=7cm]{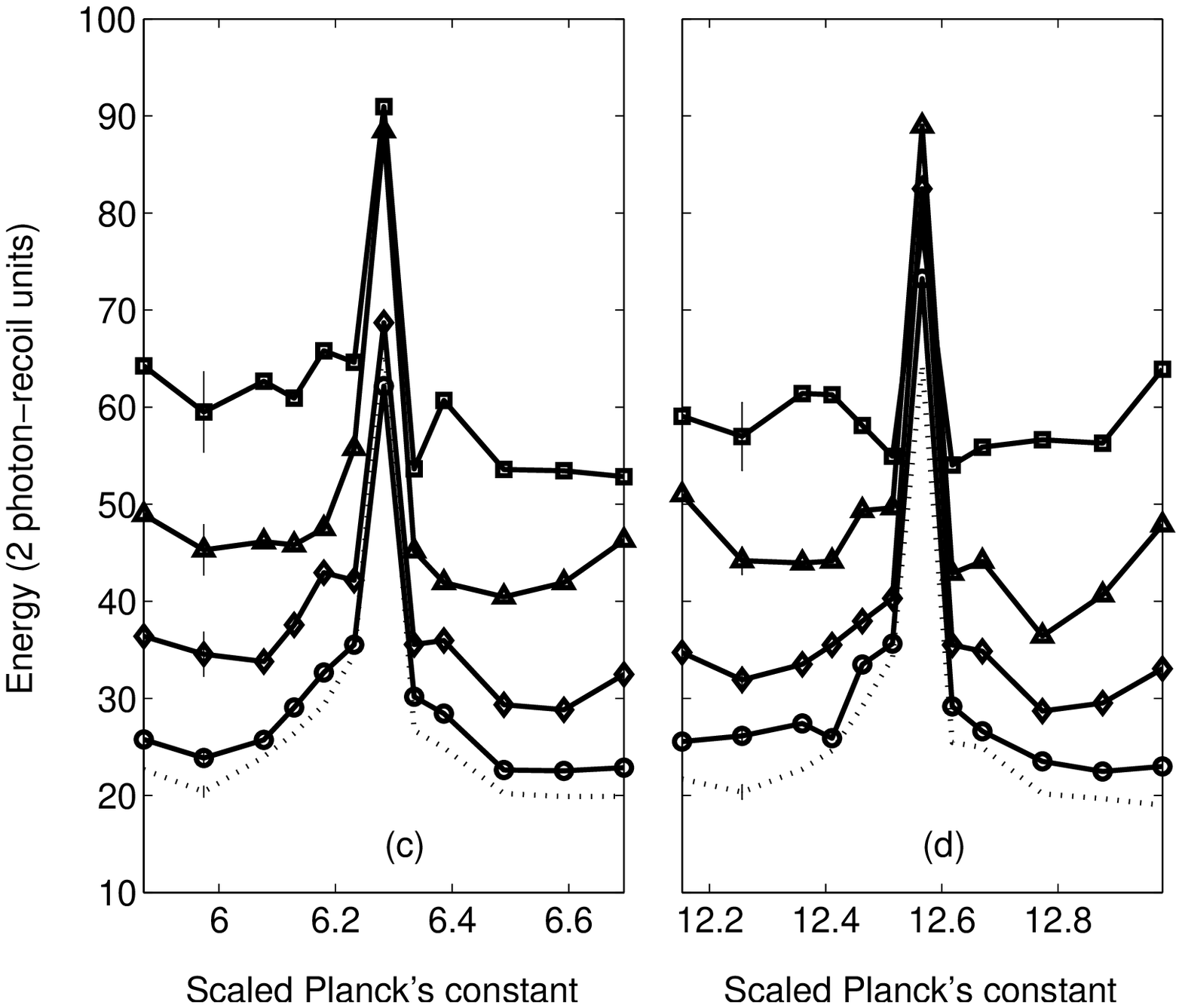}
\caption{\label{fig:qramp} Measured energies ((a) and (b)) near the first and
second primary quantum resonances and associated simulation
results ((c) and (d)) for various levels of amplitude
noise. We have taken $\kappa/\kbar=3.77$, as calculated from Eq. \ref{eq:resheightamp}
 and the spontaneous emission rate used for
simulations  is $2.5\%$. The dotted line represents the no-noise case, 
circles ${\mathcal{L}_{A}}=0.50$, diamonds ${\rm \mathcal{L}_{A}}=1.0$,
triangles ${\mathcal{L}_{A}}=1.50$ and squares ${\mathcal{L}_{ A}}=2.0$.
Sample error bars are shown on the second point for each curve.}
\end{figure}

We now present experimental measurements of the quantum resonance peaks at
$\kbar=2\pi$ and $4\pi$, in the presence of noise applied to the amplitude 
or period of the kicking pulse train(Figs. \ref{fig:qramp} and \ref{fig:qrper}).
Simulations are performed using the Monte Carlo wavefunction method as 
has previously been discussed in refs. \cite{Daley2001,Auckland2003}. 
For comparison with simulations, the value of $T$ corresponding to 
quantum resonance (that is $\kbar=2\pi$ or $4\pi$)
is taken to be the experimental position of the resonance peak.
This gives values of $T_{{\rm res},1} = 61 \mu$s and $T_{{\rm res},2} = 121.5 \mu$s which 
are within $1\%$ of the theoretical values of $2\pi/8\omega_r = 60.5 \mu$s and 
 $4\pi/8\omega_r=121 \mu$s respectively. The experimental resolution is limited by the 
spacing between consecutive values of $T$ (i.e. $0.5 \mu$s).
However, the exact position of the quantum resonances is not important to the results 
presented here which are concerned with the overall shape of the resonance peaks.

In this section, we measure the mean energy of the atomic ensemble,
which is given by $E = <\hat{p}^2>/2(2\hbar k_L)^2$. This quantity is
referred to as the energy in 2 photon--recoil units.
The height of the quantum resonance for a given number of kicks $n$ was
found in reference \cite{Wimberger2003} to be $E_{\rm res} = (1/4)(\kappa/\kbar)^2n$.
In the presence of amplitude noise, additional diffusive energy is gained which, 
for $\mathcal{L}_{A}=2$, is of size $(\kappa^2/12\kbar^2)n$ \cite{Steck2000}.
Thus, for maximal amplitude noise, 
the height of the quantum resonance energy peak is predicted to be
\begin{equation}
E_{\rm res} = \frac{1}{3} \left(\frac{\kappa}{\kbar}\right)^2 n.
\label{eq:resheightamp}
\end{equation}
We use this equation to determine the value of $\kappa$ to be used in our
amplitude noise simulations. Although this method systematically 
underestimates the true value of $\kappa$ (since small populations
of resonant atoms with high momenta cannot be detected experimentally)
it avoids the many systematic errors that arise when $\kappa$ is estimated
from power measurements of the kicking beam outside the MOT chamber. The 
values of $\kappa$ gained from this equation are consistent with those estimated
from experimental parameters.

If period noise is being applied instead, simulations show that the
energy around the second quantum resonance saturates at the quasilinear
value for the highest noise level, which is given by multiplying the
quasilinear energy growth $\kappa^2/4\kbar^2$ \cite{Rechester1980} by the
number of kicks to give
\begin{equation}
E_{\rm q.l.} = \frac{1}{4}\left(\frac{\kappa}{\kbar}\right)^2 n.
\label{eq:resheightper}
\end{equation}
Thus, having measured the height of the resonance for an amplitude noise level of $2$,
we can solve Eq. (\ref{eq:resheightamp}) for $\kappa/\kbar$ which gives $3.77 \pm 0.04$.
Similarly, having calculated the experimental quasi--linear energy of $66 \pm 0.7$ from the line
fitted in Fig. \ref{fig:qrper}(b), we can solve Eq. (\ref{eq:resheightper}) for
$\kappa/\kbar$ which gives $3.63 \pm 0.03$. Given the 
different systematic errors which arise for amplitude and period noise calculations
of $\kappa$ and the possibility of laser power drift between experimental runs, we do not 
expect perfect agreement between the two values.
Using the values of $\kappa$ gained from Eqs. (\ref{eq:resheightamp}) and (\ref{eq:resheightper})
in our simulations we find good quantitative agreement between 
experimental and simulation results. We note that period noise experiments allow $\kappa$ to be
determined more accurately because the quantum resonance behaviour is destroyed
and therefore the wings of the momentum distributions are not populated.
This leads to more accurate values for the experimentally measured 
final energies.

Once $\kappa$ has been calculated from the measured energies, the 
spontaneous emission rate per pulse may be deduced by calculating 
the Rabi frequency $\Omega$ from Eqs.(\ref{Eq:Omega_eff}) and (\ref{Eq:Omega_eff2})
and using the standard expression to find the probability of spontaneous emission
\cite{Metcalf1999}.
Measured and simulated energies are plotted against $\kbar$ (which may
also be thought of as the scaled kicking period of the kicked rotor 
system (as in \cite{Wimberger2003}) where $\kbar=2\pi$ corresponds
to the kicking period at which the first quantum resonance peak occurs).

\subsection{Amplitude noise}
In our experiments, we measured energies at pulsing periods close
to quantum resonance for the first and second quantum resonances,
which occur at $\kbar=2\pi$ and $4\pi$ respectively. Amplitude noise
was applied at levels of ${\mathcal{L}_A} = 0.5, 1, 1.5$ and $2$.
Fig. \ref{fig:qramp} shows the results obtained. 
We see that the resonance peak increases in height and that the
reduced energy level to either side of resonance rises with increasing noise level.
However, somewhat surprisingly, the resonance peak still remains 
prominent compared to the surrounding energies, 
even for the highest possible level of amplitude noise, although 
it becomes less well defined.
We note that there is essentially no difference between the behaviour 
seen at the first and second quantum resonances apart from the fact that
the energies are systematically lower for the second quantum resonance in experiments.
This is due, in part, to the fact that the atomic cloud expands to a larger 
size during kicking for the second quantum resonance as compared with the first.
This leads to a lower
average kicking strength being experienced by the atoms (a feature not included
in our simulations). Additionally, the total expansion time for the atoms,
including kicking , is constant which means that for the sweeps over the 
second quantum resonance the atoms have less free expansion time after kicking 
than at the first quantum resonance. This also leads to a systematic 
underestimation of the energy.

That the dynamics at quantum resonance itself is robust against
amplitude noise is not surprising. The resonance arises because
the time between pulses matches the condition for unity quantum 
phase accumulation after free evolution. The introduction of amplitude
noise does not affect this fundamental resonance criterion.
Seen from the point of view of atom optics, the resonant period is
the Talbot time (corresponding to $\kbar=4\pi$)
 \cite{dArcy2001, dArcy2001b}. Whilst the amplitude
of the pulses applied affects the number of atoms coupled into 
higher momentum classes, it does not affect the period dependent Talbot
effect which gives rise to the characteristic energy growth seen at 
resonance. 

The most surprising feature in these experiments is the 
survival of low energy levels to
either side of the resonance. Persistence of quantum dynamical 
localisation is the most obvious explanation for 
the sharp decrease in energy to either side of quantum resonance.
However the results of Steck et al. \cite{Steck2000} (which were performed
far from quantum resonance at $\kbar=2.08$) demonstrated 
that dynamical localisation is destroyed by high levels (corresponding to $\mathcal{L}_A = 2$) of amplitude
noise. In Section \ref{sec:EpsClass}, we will employ the recently developed $\epsilon$--classical
description of the quantum resonance peak to explain this persistence 
of localisation.

\begin{figure}[thb]
\includegraphics[height=7cm]{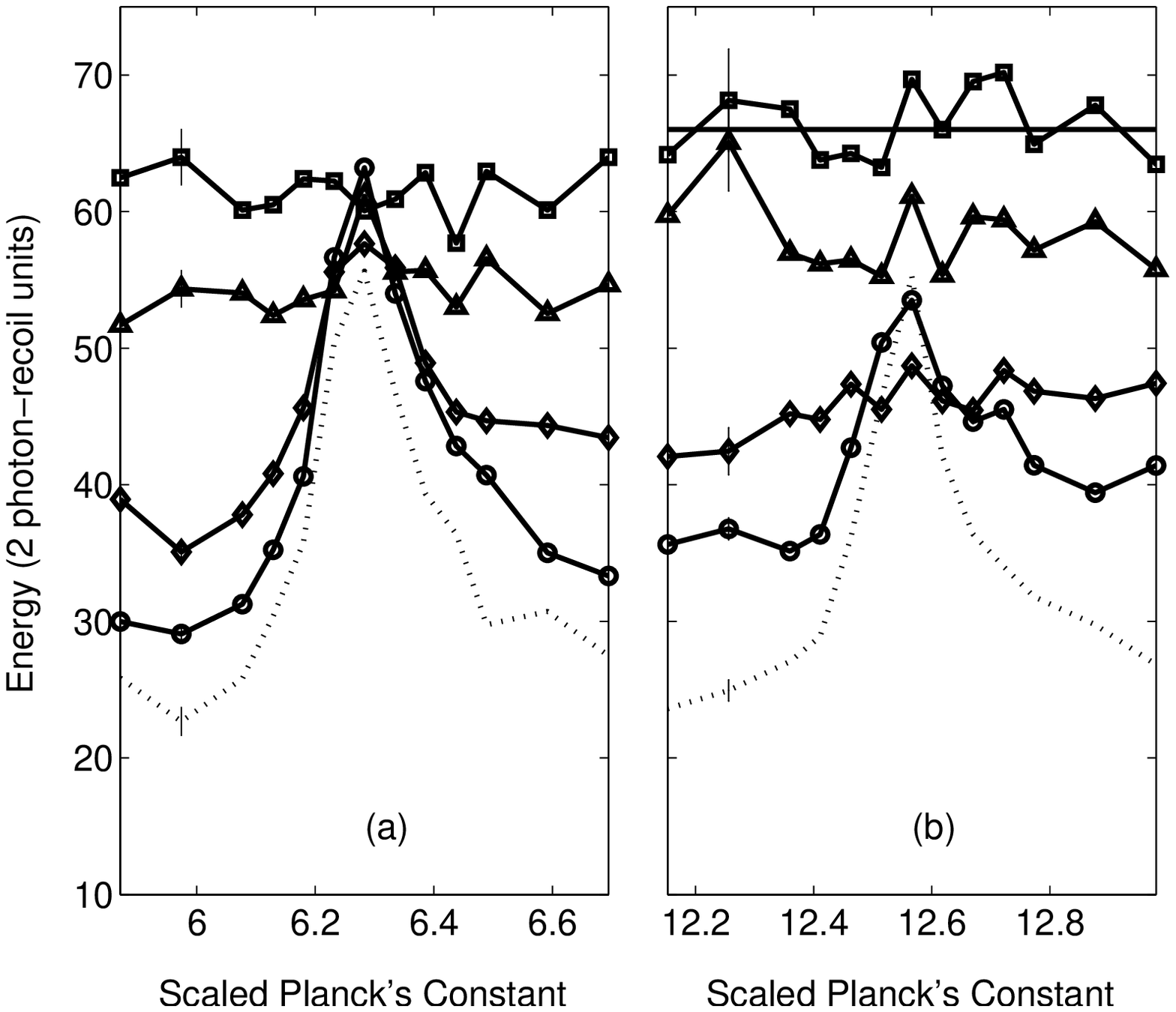}
\includegraphics[height=7cm]{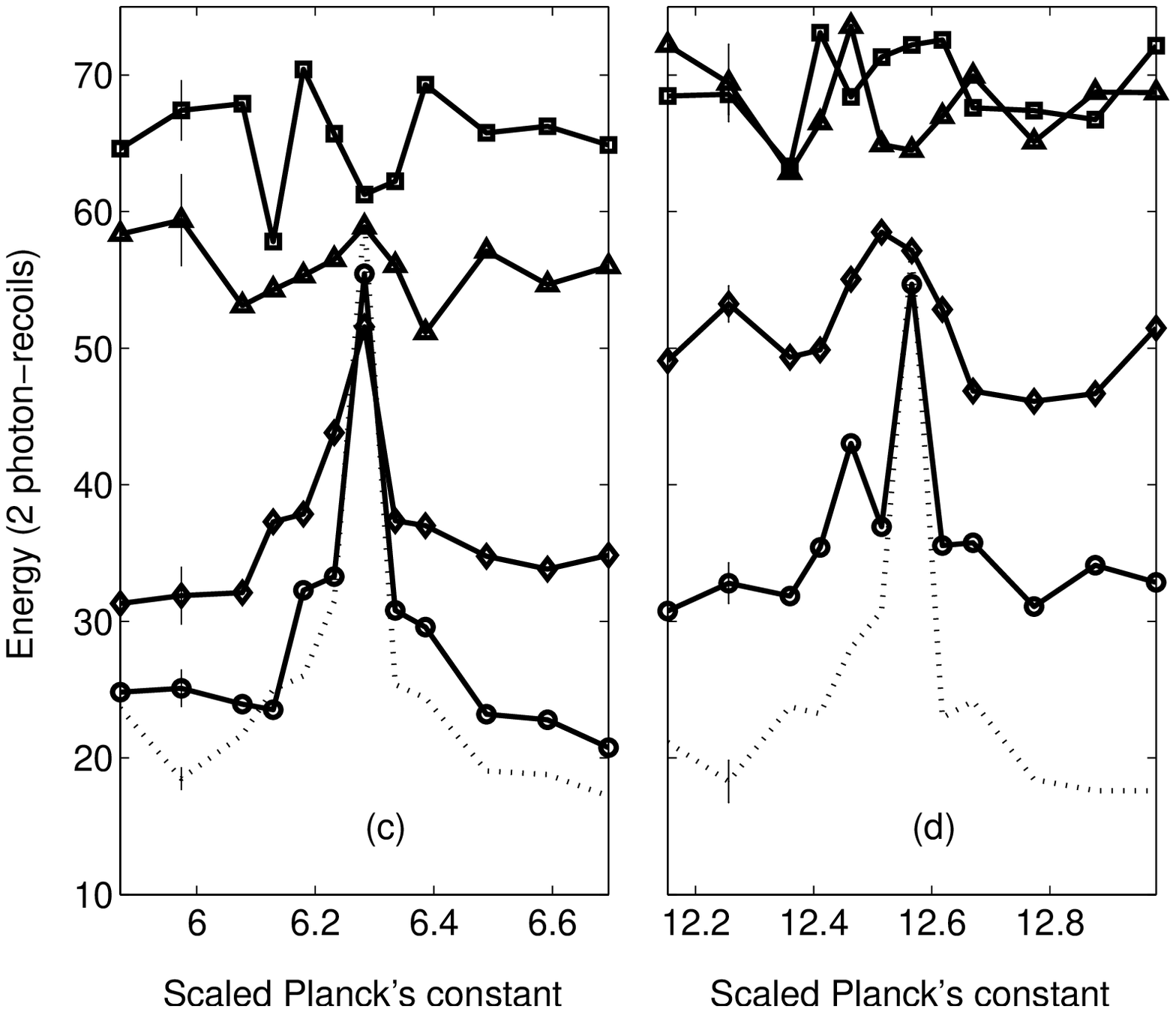}
\caption{\label{fig:qrper} Measured energies ((a) and (b)) near the first and
second primary quantum resonances and associated simulation
results ((c) and (d)) for various levels of period
noise. We have taken $\kappa/\kbar=3.61$, as calculated from Eq. \ref{eq:resheightper}
 and the spontaneous emission rate used for
simulations  is $2.5\%$ The dotted line represents the no-noise case, 
circles ${\mathcal{L}_{ P}}=0.01$, diamonds ${\mathcal{L}_{P}}=0.02$,
 triangles ${ \mathcal{L}_{ P}}=0.05$ and squares ${\mathcal{L}_{P}}=0.1$.
Sample error bars are shown on the second point for each curve.
In (b), a straight line was fitted to the energies in the highest noise case.
It gives the quasi--linear energy as $66 \pm 0.7$.}
\end{figure}

We see that the experimentally measured resonance peaks are broader than
those predicted by simulations. The broadening may result from a higher
than expected spontaneous emission rate, resulting from a small amount of
leaked molasses light which is inevitably present during the kicking cycle.
Additionally, phase jitter on the optical standing wave can be caused by
frequency instability of the kicking laser and mechanical vibrations of 
the retroreflecting mirror. Such phase noise is equivalent to a constant
level of period noise and would also lead to broadening of the resonance.
It is hard to quantify the amount of phase noise present, although the clear 
visibility of the resonances when no extra period noise is applied (see dotted line,
Fig. \ref{fig:qrper})
suggests that it is small in amplitude. However, these uncertainties do 
not affect the observation of the qualitative shape of the resonance structure under
 the application of amplitude noise and in particular, the puzzling robustness
of the low energy levels to either side of resonance.

\subsection{Period noise}
For comparison, we also present results showing the effect of period 
noise on the first two primary quantum resonance peaks.
It may be seen that even small amounts of this type of 
noise have a large effect on the near resonant dynamics.
Fig.~\ref{fig:qrper} shows the results for noise levels
of $0.01$, $0.02$, $0.05$ and $0.1$. The first primary 
quantum resonance peak is found to be very sensitive to 
small deviations from strict periodicity of the pulse train. 
Noise levels of $0.05$ and $0.1$ completely wash out the peak,
regaining the flat energy vs. kicking--period curve that we expect
in the case of zero kick--to--kick correlations.
The effect of period noise on the second primary quantum resonance 
is similar, although it is even more sensitive with an $0.02$ noise
level completely destroying the peak. At higher noise levels, the mean
energy tends towards the zero--correlation energy level.

 The greater effect of period noise on the
second quantum resonance is due to the greater absolute variation
possible in the free evolution period between pulses, since the kicking
period in this case is twice that of the first quantum resonance.
This has been verified by our group in separate experiments where the absolute
variation of the kicking period was held constant \cite{MarkThesis}. Such noise 
was found to have a more uniform effect on structures in the mean energy.
Sensitivity of the dynamics near quantum resonance to noise applied to the
kicking period is not surprising, given the precise dependence of the 
resonance phenomenon on the pulse timing. The quantum phase accumulated
between kicks is randomised and the kick--to--kick correlations destroyed.
However, the stark contrast between the sensitivity of the near resonant
dips in energy to amplitude and period noise requires further elucidation,
which we now provide by looking at the correlations which lead to quantum
resonance at early times and the $\epsilon$--classical dynamics of the kicked rotor
near quantum resonance.

\section{Reappearance of stable dynamics close to quantum resonance}
\label{sec:EpsClass}
\begin{figure}[bht]
\centering
\includegraphics[height=7cm]{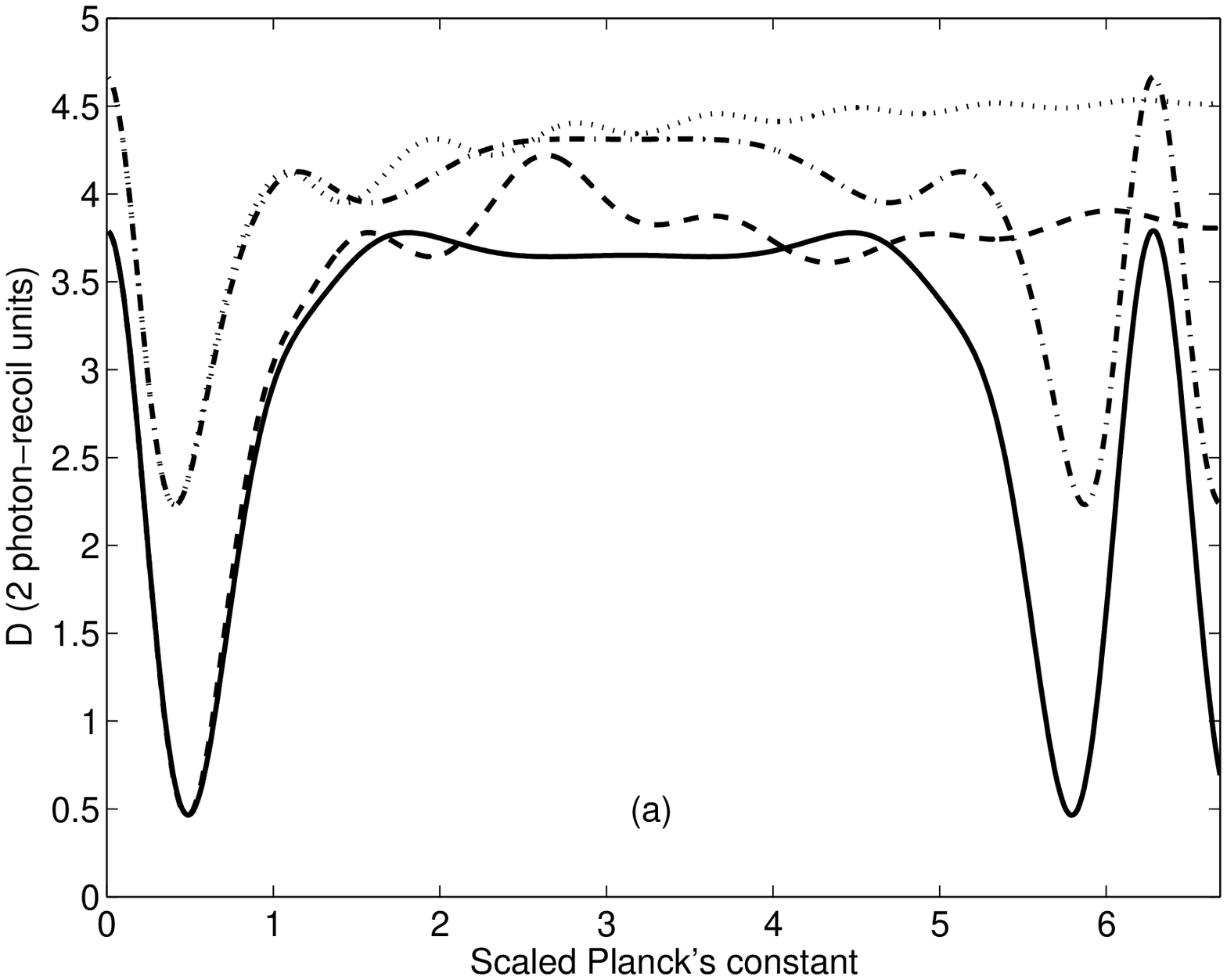}
\includegraphics[height=7cm]{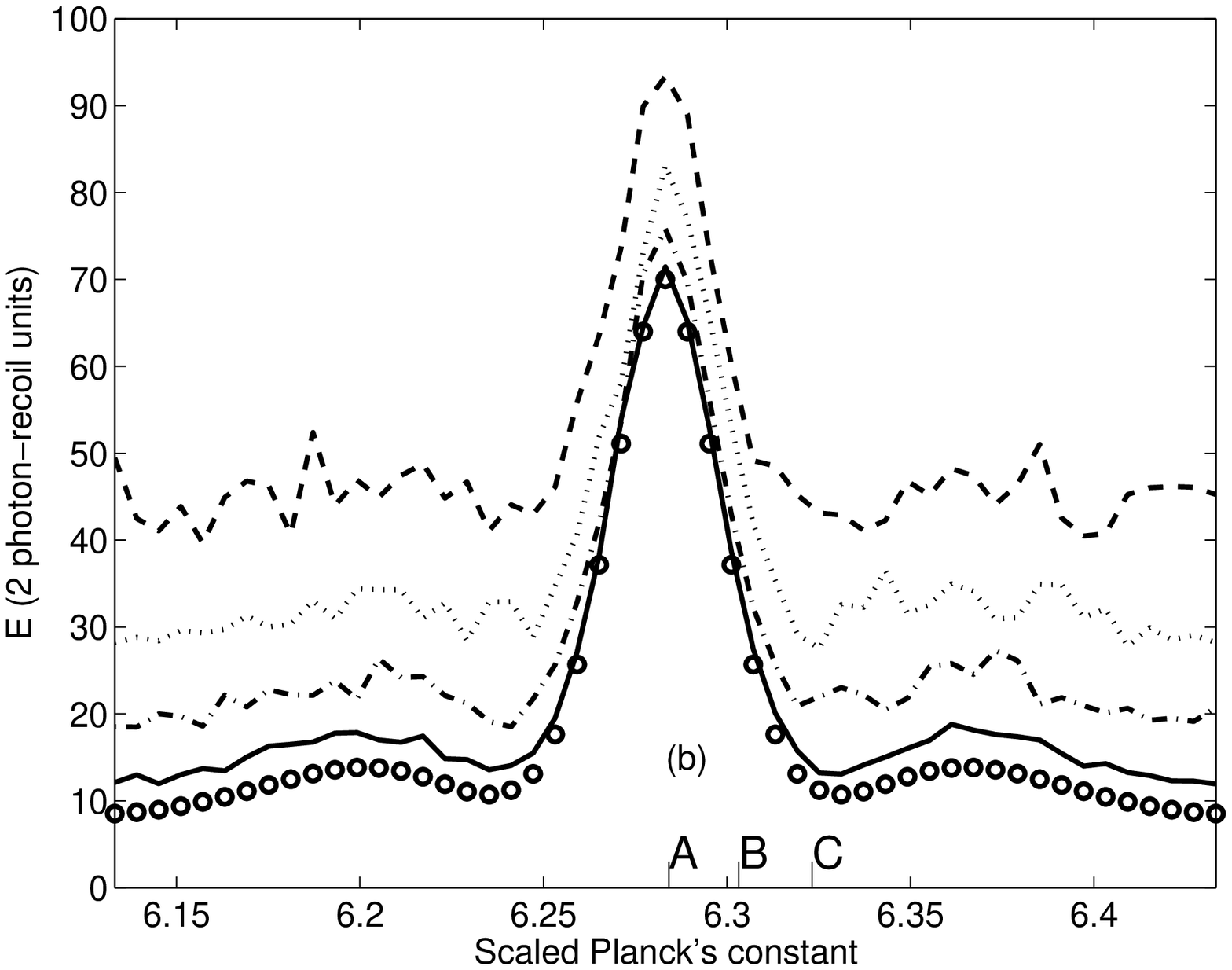}
\caption{\label{fig:early&epsclass} Analytical predictions of the effect of 
amplitude noise on the quantum resonance peak at $2\pi$ for $\kappa/\kbar =3.7$.
(a) shows the behaviour of the early energy
growth rate $D$ (as given by Eq. \ref{eq:earlyD_amp}) of the classical and quantum kicked rotor 
as a function of $\kbar$ for amplitude noise
levels of $\mathcal{L}_A = 1$ and $2$. Dashed and dotted lines are the classical 
rates for amplitude noise levels of $1$ and $2$ respectively. Solid and 
dash-dotted lines are the quantum rates for $\mathcal{L}_A=1$ and $2$
 respectively. (b) shows the resonant peak produced by 20 iterations of Eq. (\ref{eq:ecsm})
about $\kbar=2\pi$ 
with ${\mathcal{L}_A} = 0$ (circles), $0.5$ (solid line), $1.0$ (dash-dotted line), $1.5$ (dashed line)
and $2.0$ (dotted line). Each point is an average of 10 different noise realisations. The letters A--C
indicate values of $\epsilon$ as referenced from Fig. \ref{fig:phase}.}
\end{figure}

We now seek to explain the surprising resilience of the structure near quantum resonance
to the application of amplitude noise. Since the effect of amplitude noise
is the same for the resonance peaks at $\kbar = 2\pi$ and $4\pi$ we consider 
only the resonance peak about $\kbar=2\pi$, although the arguments easily generalise to
other quantum resonance peaks occuring at multiples of this value.
 We also limit our attention to the case where there
is no spontaneous emission, as this form of decoherence, at the levels present in these 
experiments, merely broadens the resonance peak and does not affect its qualitative
behaviour in the presence of amplitude noise.

The stability of the quantum resonance
structure in the late time energy (as measured in our experiments) may be 
explained by appealing to the $\epsilon$--classical 
mechanics formulated by Wimberger \emph{et al.} \cite{Wimberger2003,
Wimberger2003bpp, dArcy2003pp}. In this description of the kicked
rotor dynamics, a fictitious Planck's constant is introduced which is
referenced to zero exactly at quantum resonance. Thus, even though the quantum 
resonance peak is a purely quantum mechanical effect, its behaviour may
be well described by a (fictitious) classical map near to resonance.

Before considering this picture, however, we will look at the 
resonances found in the early time classical
and quantum energy growth rates of the kicked rotor which provide similar
 insight 
over a wider range of values for $\kbar$.
The classical rates were first derived by Rechester
and White \cite{Rechester1980} and their work was extended to
the quantum kicked rotor by Shepelyansky \cite{Shepelyansky1987}. These 
expressions for the early time classical and quantum energy growth rate, $D$, have
the advantage that they hold for any pulsing period and not just for those 
within an $\epsilon$ neighbourhood of the quantum resonance period.
Fig. \ref{fig:early&epsclass}(a) plots the early--time
energy growth rate $D$ for the classical and quantum dynamics
against the effective Planck's constant $\kbar$. For sufficiently 
large values of $\kappa/\kbar$ the energy growth rate after 5 kicks
obeys the approximate expression \cite{Shepelyansky1987}
\begin{eqnarray}
D & \approx & \frac{1}{2}\left(\frac{\kappa}{\kbar}\right)^2\left(\frac{1}{2} - J_2(K)
-J_1^2(K)\right.   \nonumber \\
& &  \left. \frac{}{} + J_2^2(K) + J_3^2(K) \right),
\label{eq:earlyD}
\end{eqnarray}
where the $J_n$ are Bessel functions and $K=\kappa$ for the classical case and
$K=\kappa_q=2\kappa\sin(\kbar/2)/\kbar$ for the quantum case. The energy growth rate is expressed in the 
same energy units used in reference \cite{Auckland2003}.
This formula was generalised by Steck \emph{et al.} \cite{Steck2000} to the case where amplitude noise
is present in the system, giving
\begin{eqnarray}
D & \approx & \frac{\kappa^2+\mathrm{Var}(\delta K)}{4\kbar^2} + \frac{\kappa^2}{2\kbar^2}\left(-\mathscr{J}_2(K)\right. \nonumber \\
 & & \left. \frac{}{} - \mathscr{J}_1^2(K) +\mathscr{J}_2^2(K)+\mathscr{J}_3^2(K)\right),
\label{eq:earlyD_amp}
\end{eqnarray}
where $K$ is defined as before, $\delta K$ is a random variable giving the fluctuation 
in $K$ at each kick, 
$\mathrm{Var}(\delta K)$ is the variance of the noise distribution $P(\delta K)$ , and 
\begin{equation}
\mathscr{J}_n(K) := \int_{-\infty}^{\infty}P(\delta K)J_n(K+\delta K){\rm{d}}(\delta K).
\label{eq:Besmod}
\end{equation}
The new functions $\mathscr{J}_n$ are averages of the normal Bessel functions over the 
noise distribution. We note that references
\cite{Steck2000} and \cite{Shepelyansky1987} deal with diffusion of the momentum $\rho$,
whereas we present our results in terms of $\rho/\kbar$. 
Hence, when comparing our results for energies or energy growth rates 
with the formulae in the aforementioned references, division by $\kbar^2$ is necessary.

Of particular interest is the behaviour 
near $\kbar=0$. We note that using Shepelyansky's formula in this regime
can be problematic because in the fully scaled system, the width of the 
initial atomic momentum distribution scales with $\kbar$ and may become
small enough that Shepelyansky's assumption of a uniform initial momentum
distribution is no longer valid \cite{Daley2002}. Assuming, however, 
that a broad initial momentum distribution may be maintained in the classical limit,
we see that a peak exists at $\kbar=0$
for both the classical and quantum dynamics and the 
classical and quantum curves match perfectly until $\kbar \sim 0.5$.
More importantly, a reduced energy region at $\kbar \approx 0.5$ remains
even for the highest level of amplitude noise, as seen in Fig.
 \ref{fig:early&epsclass}(a).
At larger values of $\kbar$, the oscillations in the classical
 growth rate are destroyed by noise. However, in the quantum case, 
the robust peak structure seen near $\kbar=0$ repeats itself at 
multiples of $\kbar = 2\pi$.
The survival of the structure near $\kbar=0$ is attributable to the
near integrability of the dynamics (classical and quantum) 
for small values of $\kbar$. We recall that in the
scaling used for these experiments the ratio $\kappa/\kbar$ is kept constant 
where $\kappa$ is the classical stochasticity parameter of the system. 
Thus we have $\kappa \rightarrow 0$ as $\kbar \rightarrow 0$.
At small values of $\kbar$ and thus $\kappa$, since the perturbation from an unkicked rotor is 
quite small, the system is near--integrable (i.e. the dynamics are stable) 
and the effect of fluctuations in the perturbation (amplitude noise)
are far less compared with the effect at higher $\kappa/\kbar$ where the system is
chaotic. Fig. \ref{fig:early&epsclass}(a) shows that, in the quantum case, this stability
reappears near quantum resonance, a fact that may be explained by inspection of Eqs. 
(\ref{eq:earlyD_amp}) and (\ref{eq:Besmod}). These equations show that the destruction of quantum 
correlations due to amplitude noise occurs due to the stochastic variation of the argument
$\kappa_q$ of the Bessel functions.  If $\kappa_q$ is small then so is the absolute variation
of $\kappa_q$ inside the Bessel functions due to amplitude noise and, therefore, there is little
damage to the quantum correlations themselves. Since $\kappa_q \rightarrow 0$ at quantum resonance,
the same behaviour seen near $\kbar=0$ reappears at integral multiples of $\kbar=2\pi$.
\begin{figure*}[bht]
\centering
\includegraphics[width=14cm]{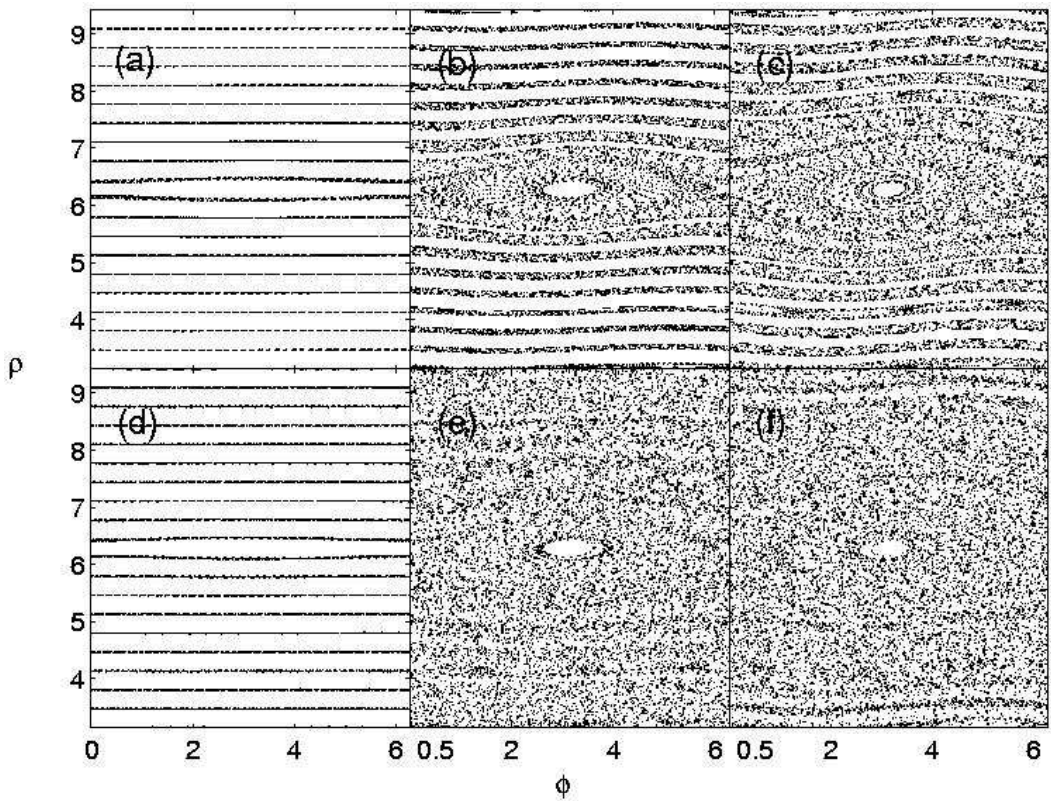}
\caption{\label{fig:phase}Phase space portraits for the 
$\epsilon$--classical standard map for $\kbar=2\pi$ and $k=3.7$.
 The figures in the first row ((a), (b) and (c)) 
are for an amplitude noise level of $0$ and for values of $\epsilon$ of $0.001$,$0.02$ and $0.04$
respectively. The second row (figures (d), (e) and (f)) 
show the $\epsilon$--classical phase space for an amplitude noise level of $2$ and the same
values of $\epsilon$. In Fig. \ref{fig:early&epsclass}(b), the values of $\kbar$ corresponding
to $\epsilon = 0.001$, $0.02$ and $0.04$ are labelled A,B and C respectively.}
\end{figure*}

The formula for the early time energy growth rate $D$ also provides us with 
predictions of the qualitative behaviour of the late time energy \cite{Auckland2003}.
However, if we limit our attention to the energies for $\kbar \approx 2\pi m$
where $m$ is a positive integer, the $\epsilon$--classical model of Wimberger
\emph{et al.} may be employed to calculate the energies around the quantum resonance
after larger numbers of kicks.
If $\epsilon=2\pi m - \kbar$ is the (small) difference between $\kbar$
and a resonant point, the dynamics of the AOKR is 
well approximated by the map \cite{Wimberger2003}
\begin{subequations}
\begin{eqnarray}
\rho_{n+1} & = & \rho_n + \tilde{k}_n\sin \phi_{n+1},\\
\phi_{n+1} & = & \phi_n + {\rm{sign}}(\epsilon) \rho_n + \pi l + \kbar\beta \mod 2\pi,
\end{eqnarray}
\label{eq:ecsm}
\end{subequations}
where $\tilde{k_n} = |\epsilon| k_n$,
 $k_n = (\kappa/\kbar)R_{A,n}$ \cite{Schlunk2003a}, $\rho_n$ and $\phi_n$
are the momentum and position at kick $n$ respectively, $\rho_0 = |\epsilon|n_0$
 for $n_0$ a positive integer and
$\epsilon = \kbar-2\pi m \ll 1$ for positive integers $m$. In this paper,
$l$ is set to $1$ without loss of generality as in references \cite{Wimberger2003,Wimberger2003bpp}.
In the reformulated dynamics, $\epsilon$ plays the part of Planck's constant and $\epsilon \rightarrow 0$
may be considered to be a quasi--classical limit.

Fig. \ref{fig:early&epsclass}(b) shows the energy peak produced by the 
$\epsilon$--classical dynamics for various amplitude noise levels. 
We see that even the maximum noise level of $2$ does not destroy
the peak, a finding that agrees with the experimental and 
simulation results presented in the previous section.
Wimberger \emph{et al.} have derived a scaling law for the ratio of the 
mean energy at a certain value of $\epsilon$ to the on--resonant energy. This ratio
is a function of $\epsilon$, the kicking strength $\kappa/\kbar$ and the kick number
\cite{Wimberger2003,Wimberger2003bpp}.
The scaling law reproduces the quantum resonance peak, and its form
is found to arise from the changes in the $\epsilon$--classical phase space
as $\epsilon$ is varied and $\kappa/\kbar$ is held constant.
These changes may be seen in Fig. \ref{fig:phase}. The first row
shows the $\epsilon$--classical phase space for increasing $\epsilon$ 
and no amplitude noise.
For the relatively unperturbed system which exists at very low
values of $\epsilon$ (say $\epsilon <0.001$), lines of constant momentum
 dominate the phase space 
(see Fig. \ref{fig:phase}(a)).

We now consider the near resonant mechanics when zero noise is present, so that
$k_n = k = {\rm constant}$ for all $n$. In this case, following Wimberger \emph{et al.},
we may calculate the kinetic energy $E_n = \epsilon^{-2}\langle \rho_n^2/2 \rangle$ 
by neglecting terms of order $\epsilon$ in Eq. (\ref{eq:ecsm}b) and iterating
Eqs. (\ref{eq:ecsm})(a) and (b) followed by averaging over $\phi_0$ and $\beta$.
Iterating Eq. (\ref{eq:ecsm}) in the limit of vanishing $\epsilon$, 
and considering for simplicity only the case where $\epsilon > 0$ and $\rho_0=0$,
we find the momentum after the $n$th kick to be \cite{Wimberger2003bpp}
\begin{equation}
\rho_n \approx \epsilon k \sum_{s=0}^{n-1}\sin(\phi_0 + \pi(1 + 2\beta)s),
\label{eq:pn}
\end{equation}
whence the mean energy may be calculated as
\begin{equation}
E_n \approx \frac{k^2}{2} \left \langle \sum_{s,s' = 0}^{n-1} \sin(\phi_0 + 
\pi(1+2\beta)s)\sin(\phi_0 + \pi(1+2\beta)s') \right \rangle,
\label{eq:En}
\end{equation}
where the average on the RHS is taken over all values of $\phi_0$.

For $\kbar=2\pi$ (as in Fig. \ref{fig:phase})
the resonant value of quasi--momentum is $\beta=0.5$ \cite{dArcy2003pp} (corresponding
to the line $\rho = 2\pi$ in the phase space figures). Substitution of this value of $\beta$
into Eq. (\ref{eq:En}) followed by averaging over a uniform distribution for 
$\phi_0$ gives $E_n \approx (k^2/2)n^2$ 
(this expression is exact when $\epsilon=0$) -- that
is ballistic growth of energy occurs at exact quantum resonance for $\beta=0.5$ --
and the mean energy of the atomic ensemble (i.e. averaging over $\beta$) is raised significantly as 
$\epsilon \rightarrow 0$ (in fact it grows linearly with kick number \cite{Wimberger2003bpp}).
Thus, the uniquely quantum energy peak found at integer multiples of $\kbar=2\pi$ may also be 
explained by a \emph{classical} resonance of the $\epsilon$--classical dynamics
 which is valid in this regime.

For larger values of $\epsilon$, the phase space of the system is significantly distorted
and the approximate expression in Eq.(\ref{eq:pn}) is no longer valid.
However, two facts in particular give a qualitative explanation for the decline in mean energy away from
exact resonance: Firstly, the most distorted area of phase space is that around $\rho=2\pi$ -- that
is the region responsible for ballistic growth for vanishing $\epsilon$ \cite{Wimberger2003}. 
Thus the number of trajectories
giving ballistic growth is drastically lessened for $\epsilon>0$. Secondly, although the phase space
region responsible for ballistic energy growth is warped, the structures which prevent stochastic 
energy growth (KAM tori) remain for $\epsilon>0$ and so the full quasi--linear rate of energy growth
is not attained. These two facts taken together give a qualitative explanation for the fall off in mean energy
as $\epsilon$ is increased, as seen in Fig. \ref{fig:early&epsclass}(b).

This qualitative explanation of the structure near quantum resonance also holds in the case where
maximal amplitude noise is applied to the system, as seen in the second row of Fig. \ref{fig:phase}. 
At exact quantum resonance ($\epsilon=0$) ballistic motion still occurs even in the presence of 
amplitude noise. For $\epsilon>0$, the phase space is distorted as before and some invariant curves 
are destroyed by the applied noise. However, even for $\epsilon=0.04$ (Fig. \ref{fig:phase}(f)),
the phase space has not become completely stochastic and so we see the same quantum resonance structure
as in the no--noise case, albeit with a lower peak--to--valley energy ratio.

The persistence of the quantum resonance structure in the presence of amplitude noise
may now be seen to be due to the reappearance of the quasi--classical
dynamics which occurs at values of $\kbar$ close to resonance value, 
and far from the actual classical limit.
The $\epsilon$--classical description which is valid in this regime is marked by a
 return to complete integrability exactly at quantum resonance.
By contrast, the extreme sensitivity of the resonant peak to even small amounts of period noise
is precisely due to the sensitivity of this approximation to the exact value of $\kbar$ (and thus
the pulse timing).
Whilst similar arguments to those used for amplitude noise might 
suggest that the resonance peak should be robust to period noise too,
it is the very reappearance of the stable dynamics which is actually
ruined by this type of noise. If the
mean deviation from periodicity is of the order of the width of the
quantum resonance peak, the suppression of energy growth to either side 
of the peak is destroyed, and the final energy approaches the 
zero correlation limit for any value of the kicking period.
Comparison with the resonance seen in the early energy growth rates in 
the actual classical limit (Fig. \ref{fig:early&epsclass}(a)) shows that the 
behaviour of the quantum resonance in the presence of amplitude noise 
is qualitatively identical to that of the classical resonance. Thus,
although the $\epsilon$--classical description of quantum resonance
employs a ``fictitious'' classical dynamics in which the effective 
Planck's constant is still far from $0$, the quantum resonance
peak may be said to mark a reappearance of classical stability in 
the kicked rotor dynamics far from the classical limit.
The experimental observation of the robustness of the 
quantum resonance peak provides a new test of the validity of
the $\epsilon$--classical model for the AOKR.
\section{Conclusion}
\label{sec:Conclusion}

We have presented experimental results demonstrating that the quantum
resonance peaks observed in Atom Optics Kicked Rotor experiments 
are surprisingly robust to noise applied to the kicking amplitude,
and that quantum resonance peaks are still experimentally detectable even
at the maximum possible noise level. 
By contrast the application of even small amounts of noise to the 
kicking period is sufficient to completely destroy the resonant peak 
and return the behaviour of the system to the zero--correlation limit.
We have shown that the
stability of the resonant dynamics in the presence of amplitude noise
is reproduced by the $\epsilon$--classical
dynamics of Wimberger \emph{et al.} Viewed in light of this theoretical
treatment, the resilience of the quantum resonance peak to  amplitude noise
is due to the reappearance of near--integrable $\epsilon$--classical dynamics 
near quantum resonance, the behaviour of which is analogous (although not identical)
to that of the kicked rotor in the actual classical limit of $\kbar \rightarrow 0$.

\section*{Acknowledgments}
The authors thank Maarten Hoogerland for his help regarding the experimental procedure.
M.S. would like to thank Andrew Daley for insightful conversations regarding
this research and for providing the original simulation programs.
This work was supported by the Royal Society of New Zealand Marsden Fund,
grant UOA016.

\end{document}